\documentclass[sigconf]{acmart}

\AtBeginDocument{%
 }

\copyrightyear{2024}
\acmYear{2024}
\setcopyright{rightsretained}
\acmConference[SIGMOD-Companion '24]{Companion of the 2024 International Conference on Management of Data}{June 9--15, 2024}{Santiago, AA, Chile}
\acmBooktitle{Companion of the 2024 International Conference on Management of Data (SIGMOD-Companion '24), June 9--15, 2024, Santiago, AA, Chile}
\acmDOI{10.1145/3626246.3654743}
\acmISBN{979-8-4007-0422-2/24/06}

\makeatletter
\gdef\@copyrightpermission{
  \begin{minipage}{0.3\columnwidth}
   \href{https://creativecommons.org/licenses/by/4.0/}{\includegraphics[width=0.90\textwidth]{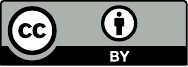}}
  \end{minipage}\hfill
  \begin{minipage}{0.7\columnwidth}
   \href{https://creativecommons.org/licenses/by/4.0/}{This work is licensed under a Creative Commons Attribution International 4.0 License.}
  \end{minipage}
  \vspace{5pt}
}
\makeatother

\usepackage{listings}
\usepackage[mathscr]{euscript}
\usepackage{amsfonts}
\begin{document}

\title{OpenIVM: a SQL-to-SQL Compiler for Incremental Computations}

\author{Ilaria Battiston}
\email{ilaria@cwi.nl}
\orcid{0009-0007-6180-364X}
\affiliation{%
  \institution{Centrum Wiskunde \& Informatica}
  \streetaddress{Science Park 123}
  \city{Amsterdam}
  \country{The Netherlands}
}

\author{Kriti Kathuria}
\email{kriti.kathuria@uwaterloo.ca}
\orcid{0009-0001-7451-6363}
\affiliation{%
  \institution{University of Waterloo}
  \streetaddress{200 University Avenue West}
  \city{Waterloo}
  \country{Canada}
}

\author{Peter Boncz}
\email{boncz@cwi.nl}
\orcid{0000-0001-6256-0140}
\affiliation{%
  \institution{Centrum Wiskunde \& Informatica}
  \streetaddress{Science Park 123}
  \city{Amsterdam}
  \country{The Netherlands}
}

\renewcommand{\shortauthors}{Ilaria Battiston, Kriti Kathuria, \& Peter Boncz}

\begin{abstract}
This demonstration presents a new Open Source SQL-to-SQL compiler for Incremental View Maintenance (IVM). 
While previous systems, such as DBToaster, implemented computational functionality for IVM in a separate system, the core principle of OpenIVM is to make use of existing SQL query processing engines and perform all IVM computations via SQL. 
This approach enables the integration of IVM in these systems without code duplication. Also, it eases its use in {\em cross-system} IVM, i.\:e. to orchestrate an HTAP system in which one (OLTP) DBMS provides insertions/updates/deletes (deltas), which are propagated using SQL into another (OLAP) DBMS, hosting materialized views.
Our system compiles view definitions into SQL to eventually propagate deltas into the table that materializes the view, following the principles of DBSP.

Under the hood, OpenIVM uses the DuckDB library to compile (parse, transform, optimize) the materialized view maintenance logic. We demonstrate OpenIVM in action (i) as the core of a DuckDB extension module that adds IVM functionality to it and (ii) powering cross-system IVM for HTAP, with PostgreSQL handling updates on base tables and DuckDB hosting materialized views on these. 
\end{abstract}



\maketitle

\section{Introduction \& Related Work}\label{intro}
Incremental View Maintenance (IVM) is the process of incrementally propagating changes in base tables into a previously computed query result instead of completely re-calculating the query from the changed base tables.

\noindent\textbf{Previous work.}
IVM has been studied in depth in the past decades, and we consider DBSP~\cite{budiu2022dbsp} the most comprehensive (and recent) work on the matter. Its core idea is to consider each query operator to be performing a stream of integrations and differentiations.
Using this principle, it then becomes possible to support arbitrarily complex queries through composition.
However, the paper that describes DBSP is conceptual and does not describe a system or strategy that implements its ideas or (consequently) any experimental performance evaluation.

The past decades have also shown that there are different approaches regarding which intermediate results to materialize and maintain.
Early strategies were conservative and did not maintain any intermediate results beyond the final materialized result table, limiting the classes of queries that could be maintained efficiently.
The DBToaster~\cite{ahmad2012dbtoaster} approach showed that an {\em aggressive} materialization strategy could provide significantly better view maintenance performance than all techniques preceding it. One should note that efficient IVM still involves trade-offs, not only because creating additional intermediates increases space overhead.
The workload also matters, namely the desired granularity and frequency of update propagation, as well as the frequency and cost of queries to the view. 
Batching changes together, for example, can amortize part of this cost but comes at the price of reduced recency.  Ideally, an IVM system should be able to apply different materialization strategies rather than a single one.

On the implementation side, many database systems offer some form of IVM; however, unlike DBSP, they typically have limitations on the complexity of the queries that can be supported.
Recent examples include the Snowflake incremental processing engine~\cite{10.1145/3589776} or Databricks Materialized Views\footnote{https://www.databricks.com/blog/introducing-materialized-views-and-streaming-tables-databricks-sql}.
Notably, both mentioned systems also assume some form of streaming: explicit SQL support for capturing updates, with Databricks specifically advocating using materialized views fed by streams of incoming tuples.

\vspace{3mm}\noindent\textbf{OpenIVM.} 
One motivation for OpenIVM is to construct pipelines that capture streams of updates in one system and feed into materialized views in another system. We call this  "\textit{cross-system}" IVM.

Our use case encompasses decentralized privacy-preserving systems: information from personal data stores flows into centralized views, while preserving privacy constraints by guaranteeing coarse-grained aggregation of sensitive attributes.\footnote{The RDDA~\cite{pub:33318} (Responsible Decentralized Data Architectures) project}.
As another example, an HTAP pipeline could be constructed with cross-system IVM by capturing deltas in an OLTP system and feeding these into an IVM computation that maintains materialized views in an OLAP system.

Previous IVM implementations, such as DBToaster~\cite{ahmad2012dbtoaster} and Materialize\footnote{https://materialize.com/}, opted to create separate computational engines.
If the goal is to deploy the IVM into existing database systems, a separate engine \textit{duplicates} query processing functionality. Following the DBSP approach, all incremental computations can be performed using relational ingredients: tables, index structures, algebraic operators, and \texttt{INSERT}/\texttt{DELETE}/\texttt{UPDATE}/\texttt{UPSERT} actions.
We, therefore, created OpenIVM as a {\em compiler}. Our implementation takes in input a database schema and view definition, and generates from there the DDL to create delta tables, possibly intermediate tables and index structures, plus SQL statements that eventually propagate updates from delta tables into the materialized view table.
The motivation for our work on SQL-to-SQL IVM is, consequently, to provide a {\em portable} framework that can be integrated into multiple database systems without adding overhead and duplicating functionality.
This utility applies either in a single system, introducing or enhancing its IVM capability, or as a bridge between two systems (cross-system IVM).

We realize that, although PostgreSQL and DuckDB use the same parser, different database systems use different SQL dialects. To ensure portability, we extend our implementation with an additional intermediate tree representation of the operators before emitting SQL, following the technique proposed in the Coral system\footnote{https://github.com/linkedin/coral}. Our approach transforms a DuckDB logical plan into a simpler abstract tree (\texttt{DuckAST}), which is then rewritten to a string in the desired SQL dialect, chosen through a flag\footnote{The authors wish to acknowledge and thank Akshat Jaimini of Thapar Institute of Engineering \& Technology for implementing this functionality.}. 

A final motivation behind OpenIVM is to provide a tool for IVM research for others to build on, allowing to experiment more easily with different intermediate result materialization strategies and other query optimizations (indexes, relational operators).

In our demonstration of OpenIVM, we show two use-cases: (i) we create a DuckDB~\cite{48-DBLP:conf/sigmod/RaasveldtM19} extension module that introduces IVM to DuckDB, powered by OpenIVM, and (ii) we show cross-system IVM with changes in PostgreSQL base tables being propagated to materialized views in DuckDB.

\section{Approach}\label{approach}

\noindent\textbf{DBSP Framework.}  The IVM problem statement is as follows: given a base table $T$ and a view $V$ defined by a query $Q$ over $T$ like $V = Q(T)$, maintain the contents of $V$ as a function of the changes to the base table. More precisely: 
let $\Delta T$ represent the change to the base table $T$ and let $\Delta V$ the change to the view $V$, as a result of executing $Q$ on the new table. Then, IVM seeks to find an algorithm $f: f(\Delta T) = \Delta V$.

DBSP presents a framework that allows finding an $f$ for any arbitrary relational query. It presents two operators $\mathscr{D}$ and $\mathcal{I}$. Let $T'$ be the table after applying a transaction (insertion, delete, update) to $T$, and $V'$ be the new view after rerunning $Q$ on the new table $T'$. The \textit{differentiation} operator $\mathscr{D}$ generates the $\Delta T$ as $T' - T$, and $\Delta V$ as $V'-V$. The \textit{integration} operator $\mathcal{I}$ performs the inverse of $\mathscr{D}$, i.e., reconstitutes the changes such as $T + \Delta T = T'$ and $V+\Delta V = V'$. The notion of $+$ and $-$ is defined precisely in the below paragraphs. DBSP presents the incremental forms of the relational operators, and a series of steps to convert a relational query $Q$ into its incremental form using the incremental operators. This incremental query takes as input $\Delta T$ and produces as output $\Delta V$. If a view, defined by a query, is $V = \pi(\sigma(T))$, then its incremental form will be $\Delta V = \pi^*(\sigma^*(\Delta T))$, where the $^*$ superscript denotes the incremental version of each operator.

OpenIVM utilizes the DBSP concepts: first, it generates the logical plan for $Q$ using the DuckDB planner. Then, rewrite rules convert the relational operators to their incremental form. Specifically, the incremental forms of selection and projection operators are the same as their relational form, and the incremental form of a join consists of three relational join operators~\cite{budiu2022dbsp}. The input to the new logical plan is the change to the base table $\Delta T$; the output is $\Delta V$, which is combined with the existing view $V$.

It is relevant to mention that the DBSP method operates on $\mathbb{Z}$-sets and not the sets consisting in the basis of relational algebra. Thus, all relational constructs defined above must be converted to $\mathbb{Z}$-sets. In order to do so, we associate a weight or \textit{multiplicity} with every element in the set, representing its frequency. For example, the $\mathbb{Z}$-set for the set \{\textit{apple}, \textit{banana}\} is \{\textit{apple} $\rightarrow$ 1, \textit{banana} $\rightarrow$ 1\}. The $+$ and $-$ in the context of the $\mathscr{D}$ and $\mathcal{I}$ operate on these weights. We use \lstinline[keywordstyle=\color{black}]{true} and \lstinline[keywordstyle=\color{black}]{false} instead of integer weights, representing respectively insertions and deletions in $\Delta T$; these multiplicities carry over to $\Delta V$. In our implementation, tuples with frequency $N$ are modeled with $N$ copies of the same element and multiplicity 1.

To combine $\Delta V$ with $V$, we perform a union for the insertions and a set difference for the deletions, following the semantics of the integration operation $\mathcal{I}$. For aggregate queries such as \texttt{SUM} and \texttt{COUNT}, the boolean multiplicity column itself is insufficient to combine the query result. Therefore, we employ the multiplicity along with the aggregation function output. For example, if the $\Delta V$ is \{\textit{apple} $\rightarrow$ (\lstinline[keywordstyle=\color{black}]{false}, 3), \textit{banana} $\rightarrow$ (\lstinline[keywordstyle=\color{black}]{true}, 1)\} and $V$ is \{apple $\rightarrow$ (\lstinline[keywordstyle=\color{black}]{true}, 5), \textit{banana} $\rightarrow$ (\lstinline[keywordstyle=\color{black}]{true}, 2)\}, then the new view $V'$ would be \{\textit{apple} $\rightarrow$ (\lstinline[keywordstyle=\color{black}]{true}, 2), \textit{banana} $\rightarrow$ (\lstinline[keywordstyle=\color{black}]{true}, 3)\}: 3 units of \textit{apple} were removed and 1 unit of \textit{banana} was added.

\begin{figure}
  \centering
  \includegraphics[width=\linewidth]{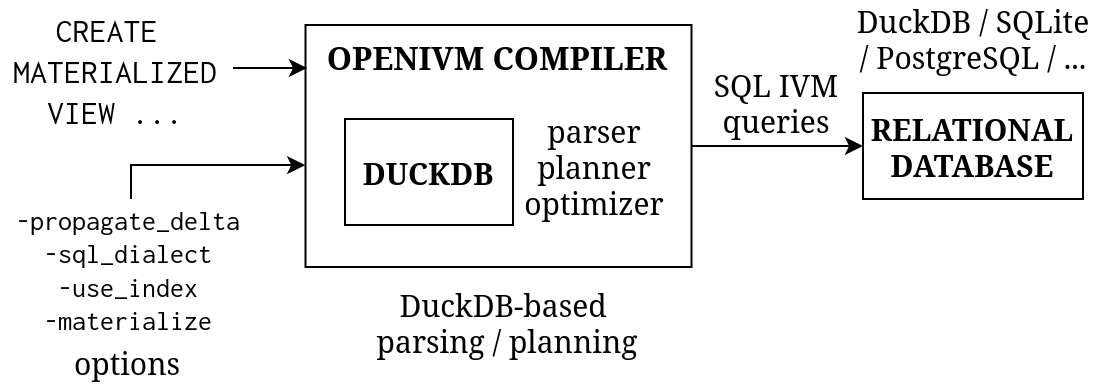}
  \vspace{-3mm}\caption{Our IVM engine implementation, consisting of a SQL-to-SQL compiler wrapped around DuckDB. Users can specify the expected optimization strategies through flags.}
  \vspace{-2mm}\label{fig:ivm}
\end{figure}

\vspace{3mm}\noindent\textbf{The Compiler: DuckDB inside OpenIVM.}
DuckDB~\cite{48-DBLP:conf/sigmod/RaasveldtM19} is an embedded analytical DBMS originally developed at CWI; it caters to many analytical use cases, including data science and data transformation pipelines, and can be deployed in mobile apps, in-browser (using WASM), as well as laptops and even cloud computing, thanks to its small footprint, portable code, and embeddable nature.

Figure \ref{fig:ivm} shows how the OpenIVM SQL-to-SQL compiler takes advantage of DuckDB being an embedded database, linking it as a library. This gives it access to the DuckDB SQL parser, planner, and optimizer, which are crucial infrastructures for IVM query rewriting and optimization.

It is not required to fork DuckDB to extend its functionality, as it allows to write {\em extension} modules in C++.
These can be loaded on the fly in a running DuckDB instance to extend its catalog, access methods, parser, optimizer, and execution operators.
Parsers, however, are notoriously non-extensible. The DuckDB approach here is first to use its own parser, but on syntax errors, try to re-parse a SQL statement with fall-back parsers provided by extensions.
An extension module registers its new functionality by calling DuckDB registration functions.
These registration functions can also be called directly from an application that uses DuckDB as a library.
This is what our SQL-to-SQL compiler does: it registers some extension functionality, namely a parser extension and extra rewrite rules in the optimizer. 

Similar to DuckPGQ~\cite{pub:32773} (which adds SQL/PGQ syntax), we developed a simple fall-back parser that recognizes the \texttt{CREATE MATERIALIZED VIEW} syntax, removes the \texttt{MATERIALIZED} keyword and finally returns into the original DuckDB parser.
The query is then transformed by DuckDB into a tree of parsed statements, then into a logical algebra tree, and subsequently optimized. 
As a final step in the optimization, DuckDB will call the OpenIVM extension rules.
Here, we substitute bindings at the leaves such that the query is executed against the changes rather than the original table. The DBSP-style rewriting of the relational query operators is performed in a bottom-up fashion, rewriting all relational operators into their incremental equivalents. 

In terms of query optimization, aggregation, for instance, allows building an index on the materialized aggregation table (using the \texttt{GROUP BY} columns as keys). Similarly, one can think of various relational strategies or custom operators to incorporate changes in a materialized aggregation: replacing the materialized table with a \texttt{UNION} and regrouping, or through a full-outer-join, or maintaining it with a left-join with an \texttt{UPSERT} (ignoring deletions here for simplicity).
The materialization strategy (eager, none, or something in between) provides a similar search space for alternative IVM computation plans. 
For now, we only offer a small number of alternatives, and choosing one is controlled manually using compiler switches; but as we implement join operations, the search space should increase, and \textit{cost-based} optimization should then make these choices, paving the way for new research opportunities. 

Another important open question in cross-system IVM is how to propagate changes from $T$ to $\Delta T$. This process could be done in many ways: through triggers, optimizer rules, or not at all in systems such as append-only streams. This choice is also currently left to the user -- optimizer rules are being developed for DuckDB; however, for PostgreSQL (or any alternative system), users are required to configure these triggers independently. 

\begin{figure}
  \centering
  \includegraphics[width=\linewidth]{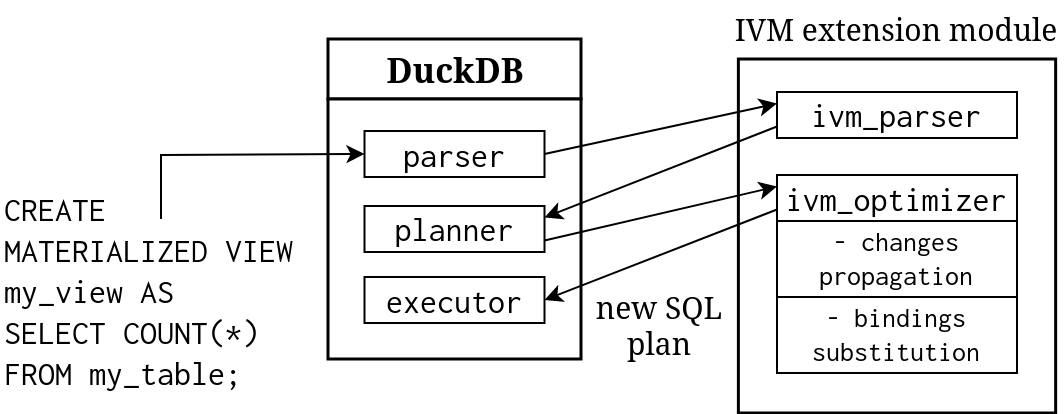}
  \vspace{-3mm}\caption{Our IVM optimizer extension, interacting with the DuckDB core engine.}
  \vspace{-2mm}\label{fig:duckdb}
\end{figure}

\vspace{3mm}\noindent\textbf{The Extension Module: OpenIVM inside DuckDB.}
While the OpenIVM SQL-to-SQL compiler can be used as a standalone command-line tool, we also wrap it into a dynamically loadable DuckDB extension module. Our implementation, as shown in Figure \ref{fig:duckdb}, extends the DuckDB query processing pipeline. The purpose of such, specifically, is to add IVM to DuckDB \textit{natively}.
To achieve this, when the fall-back parser parses a \texttt{CREATE MATERIALIZED VIEW}, we execute the compiled output to create the delta tables as well as any generated intermediate result tables or indexes, along with a table that represents the materialized result. 
We store the SQL scripts that propagate the contents of the delta tables to the materialized view table on the disk to allow future inspection and usage without having to start DuckDB.
Furthermore, another optimizer rule can then be used to intercept \texttt{INSERT}/\texttt{DELETE}/\texttt{UPDATE} statements into the base tables. These are translated into actions that carry through these base table changes, fill the delta tables $\Delta T$, and kick off the SQL propagation scripts.

One single materialized view $V$ definition corresponds to multiple SQL instructions: the $\Delta T$ are created for each $T$, along with their eventual indexes and boolean multiplicity columns. Furthermore, the $\Delta_i V$ are generated, where $i = 1, \dots N$ and $N$ is the number of views to materialize. Currently, our IVM implementation only supports single-table ($N = 1$) projections, filters, grouping, and the aggregates \texttt{SUM} and \texttt{COUNT}; we are in the process of extending it to \texttt{MIN}, \texttt{MAX} and \texttt{JOIN}.

Internally, we store materialized views as tables and save their additional properties -- query plan, SQL string, query type -- in metadata tables. Once the setup for IVM has been generated, we can process the queries to compute incrementally. This step happens by implicitly calling a table function, adding a dummy node to the plan of the original query to trigger the optimizer rules. 

However, more is needed to translate the newly generated logical representation to SQL: IVM requires multiple post-processing steps to ensure consistency of $\Delta T$. These comprehend:
\begin{enumerate}
\item Insertion in $\Delta V$ of the tuples resulting from querying $\Delta T$.
\item Insertion or update in $V$ of the newly-inserted tuples in $\Delta V$, removing the multiplicity column.
\item Deletion of the invalid rows in $V$, e.\:g. the ones with \texttt{SUM} or \texttt{COUNT} equal to 0, or \lstinline[keywordstyle=\color{black}]{false} multiplicity without aggregate.
\item Deletion from $\Delta T$ and $\Delta V$ after applying the changes.
\end{enumerate}
The SQL instructions emitted for Step 2 can drastically change depending on the input query. For example, a \texttt{GROUP BY} query with an aggregation function can be translated into a \texttt{LEFT JOIN} rather than a \texttt{UNION}, depending on the complexity of operators. 

As we show during the demonstration, preliminary results indicate clear improvements in resource consumption by executing incremental computations rather than running the query against the whole dataset. We also examine the overhead caused by having a materialized index: its creation is necessary for aggregate queries, as DuckDB requires an index to apply upserts. 
The ART (Adaptive Radix Tree) is generated after having populated $V$, as it is more efficient to build small indexes for each chunk and merge them. However, its creation only adds significant overhead the first time, and it can be used in the future to speed up joins, especially when the joined part has just a few unique keys. 


\section{Demonstration Scenarios}\label{demo}
\vspace{0mm}\noindent\textbf{Enabling IVM inside DuckDB}
We demonstrate our system, providing visitors with two scenarios. We set up a GitHub repository containing our OpenIVM extension\footnote{https://github.com/ila/duckdb/tree/rdda/extension/openivm} that implements table functions to run queries and benchmarks. Our demonstration aims to familiarize users with incremental representations of computations in SQL and how easily they can be plugged into relational systems. 

We allow visitors to run arbitrary queries on our DuckDB OpenIVM infrastructure. Furthermore, we provide pre-loaded datasets to experiment with and additional ideas for queries to write in the DuckDB shell. Users can then examine the compiled output and check the correct result of the incremental computations applied to $V$. Here, we show an example of what a query emitted by our compiler looks like and the necessary steps to obtain it. 

\begin{lstlisting}[caption={Example DDL for our IVM setup}, label={lst:ddl}]
CREATE TABLE groups(group_index VARCHAR, group_value INTEGER);
CREATE MATERIALIZED VIEW query_groups AS SELECT group_index, SUM(group_value) AS total_value FROM groups GROUP BY group_index;
\end{lstlisting}

Given this DDL, our IVM compiler will emit new queries (omitted for brevity) to provide the required tables $\Delta T$ and $\Delta V$ to store changes. Then, the compiler generates the SQL statements to propagate modifications, as shown below.

\begin{lstlisting}[caption={Generated SQL instructions for IVM}, label={lst:ivm}]
INSERT INTO delta_query_groups
SELECT group_index, SUM(group_value) AS total_value, _duckdb_ivm_multiplicity
FROM delta_groups
GROUP BY group_index, _duckdb_ivm_multiplicity;
INSERT OR REPLACE INTO query_groups
WITH ivm_cte AS (
SELECT group_index, 
    SUM(CASE WHEN _duckdb_ivm_multiplicity = FALSE THEN -total_value ELSE total_value END) AS total_value
FROM delta_query_groups
GROUP BY group_index)
SELECT query_groups.group_index, 
    SUM(COALESCE(query_groups.total_value, 0) + delta_query_groups.total_value)
FROM ivm_cte AS delta_query_groups
LEFT JOIN query_groups ON query_groups.group_index = delta_query_groups.group_index
GROUP BY query_groups.group_index;
DELETE FROM query_groups WHERE total_value = 0;
DELETE FROM delta_query_groups;
\end{lstlisting}
These SQL commands can either be run eagerly, i.\:e. every time a change is registered on the base table, or lazily, i.\:e. refreshing the materialized view when it is queried. In our examples, we choose to employ the latter approach.

Once we demonstrate the potentialities of our compiler, we want to give users the possibility to benchmark it and test it in DuckDB. We offer different benchmarks with sets of pre-written \texttt{GROUP BY} queries to show how computationally intensive each part of the incremental maintenance is. We argue that the incremental computation approach is more efficient than recalculating $V$ each time it is queried. 

\begin{figure}
  \centering
  \includegraphics[width=\linewidth]{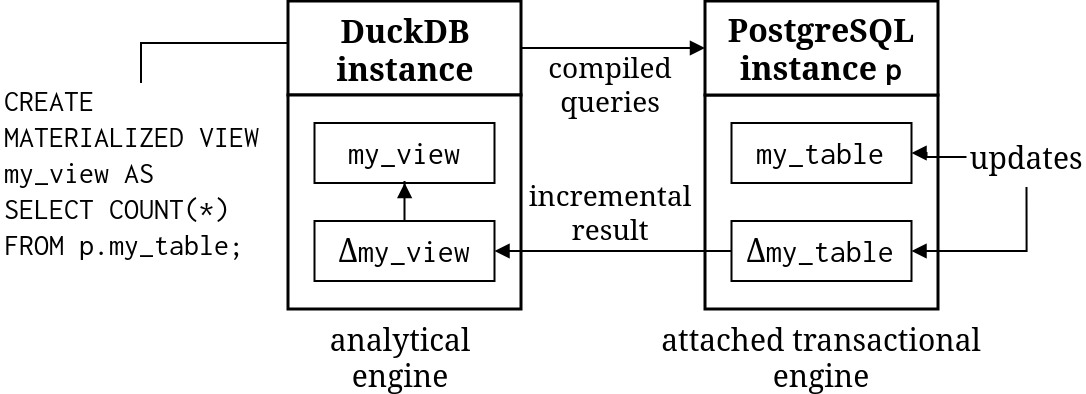}
   \vspace{-3mm}\caption{Our cross-system IVM demo, showcasing an HTAP workload with DuckDB and PostgreSQL.}
   \vspace{-2mm}\label{fig:cross_system}
\end{figure}

\vspace{3mm}\noindent\textbf{Cross-System IVM}
The architecture for our cross-system IVM is depicted in Figure \ref{fig:cross_system}: we emulate a transactional workload with PostgreSQL using the datasets previously mentioned. We allow users to input analytical queries to be run from DuckDB to an attached PostgreSQL instance. The data stored on PostgreSQL is accessed via the DuckDB integration with PostgreSQL, which gives access to the PostgreSQL schemata and allows cross-system queries.
The result of the analytical query is stored in a materialized view in DuckDB. This output is incrementally maintained using the SQL plans generated by our IVM compiler. Visitors can query data stored on PostgreSQL using the DuckDB shell. Furthermore, we provide a script to run the pipeline automatically given an input query $Q$. The output will be the result of $Q$ on the data on PostgreSQL and its changes, stored in a DuckDB materialized view.

We also allow users to benchmark our system: we show a transparent comparison of the query performance in pure DuckDB, pure PostgreSQL, cross-system, and without IVM. 

Using the OpenIVM system, therefore, makes it possible to combine a trusted and efficient OLTP system (PostgreSQL) with an efficient analytical engine (DuckDB) and also easily transform data from operational tables that are maintained using OLTP into warehoused views for OLAP use cases; thus providing a practical HTAP approach.
\bibliographystyle{ACM-Reference-Format}
\bibliography{sample-base}

\end{document}